\PassOptionsToPackage{unicode}{hyperref}
\PassOptionsToPackage{hyphens}{url}
\documentclass[
]{article}
\usepackage{multirow}
\usepackage{amsmath,amssymb}
\usepackage{lmodern}
\usepackage{iftex}
\ifPDFTeX
  \usepackage[T1]{fontenc}
  \usepackage[utf8]{inputenc}
  \usepackage{textcomp} 
\else 
  \usepackage{unicode-math}
  \defaultfontfeatures{Scale=MatchLowercase}
  \defaultfontfeatures[\rmfamily]{Ligatures=TeX,Scale=1}
\fi
\IfFileExists{upquote.sty}{\usepackage{upquote}}{}
\IfFileExists{microtype.sty}{
  \usepackage[]{microtype}
  \UseMicrotypeSet[protrusion]{basicmath} 
}{}
\makeatletter
\@ifundefined{KOMAClassName}{
  \IfFileExists{parskip.sty}{%
    \usepackage{parskip}
  }{
    \setlength{\parindent}{0pt}
    \setlength{\parskip}{6pt plus 2pt minus 1pt}}
}{
  \KOMAoptions{parskip=half}}
\makeatother
\usepackage{xcolor}
\usepackage{longtable,booktabs,array}
\usepackage{calc} 
\usepackage{etoolbox}
\makeatletter
\patchcmd\longtable{\par}{\if@noskipsec\mbox{}\fi\par}{}{}
\makeatother
\IfFileExists{footnotehyper.sty}{\usepackage{footnotehyper}}{\usepackage{footnote}}
\makesavenoteenv{longtable}
\usepackage{graphicx}
\makeatletter
\def\maxwidth{\ifdim\Gin@nat@width>\linewidth\linewidth\else\Gin@nat@width\fi}
\def\maxheight{\ifdim\Gin@nat@height>\textheight\textheight\else\Gin@nat@height\fi}
\makeatother
\setkeys{Gin}{width=\maxwidth,height=\maxheight,keepaspectratio}
\makeatletter
\def\fps@figure{htbp}
\makeatother
\newlength{\cslhangindent}
\newlength{\csllabelwidth}
\newlength{\cslentryspacingunit} 
\newenvironment{CSLReferences}[2] 
 {
  \setlength{\parindent}{0pt}
  \ifodd #1
  \let\oldpar\par
  \def\par{\hangindent=\cslhangindent\oldpar}
  \fi
  \setlength{\parskip}{#2\cslentryspacingunit}
 }%
 {}
\usepackage{calc}

\ifLuaTeX
  \usepackage{selnolig}  
\fi
\IfFileExists{bookmark.sty}{\usepackage{bookmark}}{\usepackage{hyperref}}
\IfFileExists{xurl.sty}{\usepackage{xurl}}{} 
\hypersetup{
  pdftitle={Exploring Latent Spaces of Tonal Music using Variational Autoencoders},
  pdfauthor={Nádia Carvalho, Gilberto Bernardes},
  hidelinks,
  pdfcreator={LaTeX via pandoc}}

\title{Exploring Latent Spaces of Tonal Music using Variational
Autoencoders}
\author{
Nádia Carvalho, Gilberto Bernardes \\ \small Faculty of Engineering of University of Porto, INESC-TEC \\ \footnotesize up201208223@up.pt  }

\date{}

\begin{document}

\maketitle
\begin{abstract}
Variational Autoencoders (VAEs) have proven to be effective models for
producing latent representations of cognitive and semantic value. We
assess the degree to which VAEs trained on a prototypical tonal music
corpus of 371 Bach\textquotesingle s chorales define latent spaces
representative of the circle of fifths and the hierarchical relation of
each key component pitch as drawn in music cognition. In detail, we
compare the latent space of different VAE corpus encodings --- Piano
roll, MIDI, ABC, Tonnetz, DFT of pitch, and pitch class distributions
--- in providing a pitch space for key relations that align with
cognitive distances. We evaluate the model performance of these
encodings using objective metrics to capture accuracy, mean square error
(MSE), KL-divergence, and computational cost. The ABC encoding performs
the best in reconstructing the original data, while the Pitch DFT seems
to capture more information from the latent space. Furthermore, an
objective evaluation of 12 major or minor transpositions per piece is
adopted to quantify the alignment of 1) intra- and inter-segment
distances per key and~ 2) the key distances to cognitive pitch spaces.
Our results show that Pitch DFT VAE latent spaces align best with
cognitive spaces and provide a common-tone space where overlapping
objects within a key are fuzzy clusters, which impose a well-defined
order of structural significance or stability --- i.e., a tonal
hierarchy. Tonal hierarchies of different keys can be used to measure
key distances and the relationships of their in-key components at
multiple hierarchies (e.g., notes and chords). The implementation of our
VAE and the encodings framework are made available online.
\end{abstract}

\textbf{keywords: }Symbolic Musical Encodings, Latent Spaces,
Variational Autoencoders

\hypertarget{introduction}{%
\section{1. Introduction}\label{introduction}}

One promising avenue in music cognition is using artificial neural
networks to produce latent (or embedding) representations of musical
data (\protect\hyperlink{ref-temp_id_9699009762824167}{Kim, 2022};
\protect\hyperlink{ref-temp_id_08049428684638582}{Qiu, Li, \& Sung,
2021}). In particular, Variational Autoencoders (VAEs) have shown great
potential in generating meaningful and interpretable latent spaces
(\protect\hyperlink{ref-https:ux2fux2fdoi.orgux2f10.48550ux2farxiv.1803.05428}{Roberts
et al., 2018};
\protect\hyperlink{ref-https:ux2fux2fdoi.orgux2f10.48550ux2farxiv.2212.00973}{Guo,
Kang, \& Herremans, 2022}). Latent spaces are a mathematical
representation that allows for manipulating and analyzing data using
machine learning techniques and have been successfully adopted in
various applications, such as music recommendation systems, style
transfer, and music generation. They have shown promising results in
improving the quality of generated music
(\protect\hyperlink{ref-https:ux2fux2fdoi.orgux2f10.48550ux2farxiv.1803.05428}{Roberts
et al., 2018}; \protect\hyperlink{ref-temp_id_7657716331815574}{Turker,
Dirik, \& Yanardag, 2022};
\protect\hyperlink{ref-bryan-kinns2021exploring}{Bryan-Kinns et al.,
2021}; \protect\hyperlink{ref-temp_id_7741585219955907}{Mezza, Zanoni,
\& Sarti, 2023}).

\begin{figure}[h]
\centering
\includegraphics[width=0.65\textwidth]{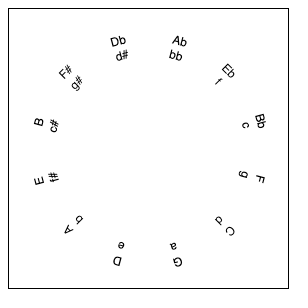}
\caption{Dimensions 1 and 2 of the four-dimensional multidimensional scaling
solution of the intercorrelations between the 24 major and minor key
profiles (Circle of fifths), as presented by \protect\hyperlink{ref-krumhansl:1990}{Krumhansl (1990)} from
probe-tone experiments.}
\label{fig:figure-1}
\end{figure}

Symbolic music is typically represented as a temporal sequence of
discrete symbols. By computing its latent space, symbolic music can be
represented as a \textbf{continuous} geometrical space, where
multidimensional vectors represent a symbol. Geometric spaces of musical
symbols have two main appeals. First, the quality of a musical symbol
depends on its spatial relationship with other symbols, i.e., its
configurable properties. Its hierarchical dependencies are typically
shown when projecting symbols segmented on unitary pitch structures,
such as notes and chords. Second, geometrical spaces of low
dimensionality provide a concise summary of relations in a form that is
easy to visualize and intuitive to understand (i.e., the \textbf{circle
of fifths} in Figure~\ref{fig:figure-1}. The
continuity of the space allows mathematical operations to be performed
on the symbolic data, such as similarity comparisons or clustering,
making it easier for machine learning algorithms to learn patterns and
generate new music.~

The intelligibility and high explanatory power of tonal pitch spaces
usually account for a variety of subjective and contextual factors.
Historically, tonal spaces can be roughly divided into two categories,
each anchored to a specific discipline and applied methods. We have
models grounded in music theory
(\protect\hyperlink{ref-temp_id_043057691023887346}{Cohn, 1997},
\protect\hyperlink{ref-temp_id_22337014670269895}{1998};
\protect\hyperlink{ref-lewin2010generalized}{Lewin, 1987};
\protect\hyperlink{ref-tymoczko:10}{Tymoczko, 2010};
\protect\hyperlink{ref-weber1832versuch}{Weber, 1817-1821}), and models
based on cognitive psychology
(\protect\hyperlink{ref-krumhansl:1990}{Krumhansl, 1990};
\protect\hyperlink{ref-longuet:87}{Longuet-Higgins, 1987};
\protect\hyperlink{ref-temp_id_38492172090114884}{Shepard 1982}).
Tonal pitch spaces based on music theory rely on musical knowledge,
experience, and the ability to imagine complex musical structures to
explain which structures work. Cognitive psychology intends to capture
and assess the mental processes underlying and relating musical pitch
from listening experiments. ~Despite their inherent methodological
differences, they share the same motivation to capture intuitions about
the closeness of tonal pitch, which is an important aspect of our
experience of tonal music
(\protect\hyperlink{ref-temp_id_6497209108989883}{Deutsch, 1984}) and
allow the quantification of pitch relations as distances (e.g., What
pitch E or G is closer to the A major key?).

Recently, data-driven approaches to the construction of pitch spaces
have been pursued from large datasets of symbolic music.
\protect\hyperlink{ref-temp_id_4779845625814849}{Moss, Neuwirth, \&
Rohrmeier (2022)} explore fundamental tonal relations in musical
compositions from a corpus representative of historical periods with the
aim of studying the evolution of tonal relations across history.
\protect\hyperlink{ref-RePEc:wsi:acsxxx:v:25:y:2022:i:05n06:n:s0219525922400082}{Nardelli,
Culbreth, \& Fuentes (2022)} propose a dynamical score network to
represent harmonic progressions from an extensive musical corpus
spanning 500 years of Western classical music. They found increased
harmonic complexity over the historical evolution of the corpora.
\protect\hyperlink{ref-https:ux2fux2fdoi.orgux2f10.5281ux2fzenodo.4285410}{Plitsis
et al. (2020)} and
\protect\hyperlink{ref-https:ux2fux2fdoi.orgux2f10.48550ux2farxiv.2109.03454}{Prang
\& Esling (2021)} explore a large corpus of monophonic and polyphonic
musical data, respectively, to evaluate the use of symbolic music
encodings in generative models. The former adopts a simple Long-Short
Term Memory (LSTM) structure, for which the ABC notation presented the
best results overall. The latter adopts the MusicVAE architecture, for
which a signal-like representation reflects better reconstruction
performance and a latent space more aligned with cognitive musical
qualities.~Our paper is in line with both works, particularly the last
one, but our proposed encoding is simpler than theirs.

The choice of musical encoding is fundamental to the performance of
machine and deep-learning techniques in symbolic music tasks, such as
generation, transcription, and style recognition
(\protect\hyperlink{ref-https:ux2fux2fdoi.orgux2f10.48550ux2farxiv.2302.05393}{Sarmento
et al., 2023}). The properties of the encoding determine the amount and
quality of information that can be extracted from the data and thus
influence the accuracy and expressiveness of the generated output. For
instance, selecting an encoding that can capture high-level semantic
features of the data, such as chord progressions or melody patterns, can
potentially improve the musical output
(\protect\hyperlink{ref-https:ux2fux2fdoi.orgux2f10.48550ux2farxiv.2302.05393}{Sarmento
et al., 2023}). Our paper explores the effectiveness of VAEs while
conditioning the model and its ability to produce latent spaces that
represent the cognitive pitch distances. To this end, we train VAEs on a
prototypical tonal music corpus of 371 Johann Sebastian
Bach\textquotesingle s (JSB) chorales and compare the latent space of
typical VAE corpus encodings --- piano roll, MIDI, ABC, Tonnetz, DFT of
pitch, and pitch class distributions. The two latter encodings are
proposed in this article and aim to leverage the potential of the DFT of
pitch and pitch class distributions in exposing higher-level information
on interval and pitch content
(\protect\hyperlink{ref-temp_id_851349609824851}{Amiot 2016}).~

Our evaluation adopts four-fold objective metrics to evaluate the
models\textquotesingle{} performance: accuracy, mean squared error
(MSE), Kullback--Leibler divergence (KL-divergence), and the
computational cost (of each encoding, training the respective VAE, and
extracting the information from the original source). Moreover, we
quantify the degree to which the encoding's latent spaces provide a
pitch space for key relations that align with cognitive distances. In
detail, we adopt intra- and inter-segment distances per key and the key
distances across all 12 major or minor transpositions per piece to
assess the degree of key segmentation between keys and their alignment
to the circle of fifths. In eliciting VAE latent spaces with a
cognitive, perceptual, and musical theoretical value from tonal music
corpus, we can foresee future endeavors which leverage pitch spaces for
style-specific musical expressions or less studied harmonic systems,
such as modal and microtonal music.

The remainder of this paper is structured as follows.
\protect\hyperlink{methodology}{Section 2}
describes the methodology of our paper.
\protect\hyperlink{symbolic-music-encodings}{Section
3} presents each encoding's implementation and characteristics.
\protect\hyperlink{vae-model}{Section 4}
presents our implementation of the VAE model.
\protect\hyperlink{evaluation-and-discussion}{Section
5} describes our two-folded approach for evaluating both the model
performance and the effectiveness of latent spaces in representing
cognitive distances between musical pitches. Finally,
\protect\hyperlink{conclusions-and-future-work}{Section
6} presents the conclusions and avenues for future work.

\hypertarget{methodology}{%
\section{2. Methodology}\label{methodology}}

\begin{figure}[h]
\centering
\includegraphics[width=1\textwidth]{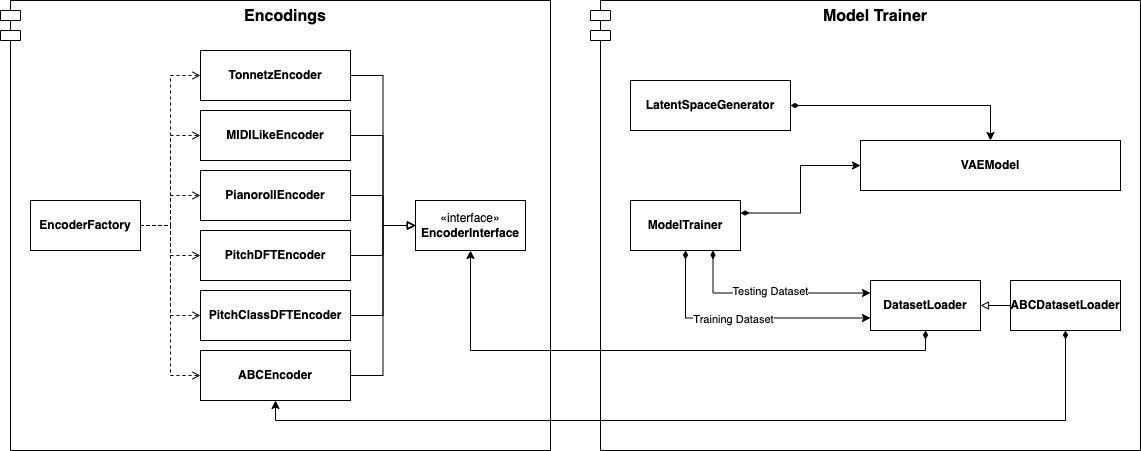}
\caption{Architecture of the proposed multi-encoding framework. It is constructed
in a modular way, allowing the training of our VAE model using the same
methods (i.e., encoding extraction, decoding of the predictions,
augmentation, one-hot encoding, storage, and retrieval of the encoded
dataset). In an application such as the one we constructed, we would
simply need to call the EncoderFactory class with the encoding's name
and the ModelTrainer class with the preferred parameters for training
the encoding's dataset.}
\label{fig:figure-2}
\end{figure}

Figure~\ref{fig:figure-2} shows the
architecture of a system we implemented to compare several symbolic
music encodings. To process encoding based on the same general
properties, we developed a Python 3\footnote{\url{https://www.python.org/,}
  Last accessed on 07/03/2023.} framework, relying on the
music21\footnote{\url{http://web.mit.edu/music21/,} Last accessed on
  07/03/2023.} library to parse music from different sources (e.g.,
MusicXML, MIDI, ABC).

For each encoding, we perform musical data augmentations by transposing
each piece to all 12 key transpositions per mode. In
\protect\hyperlink{symbolic-music-encodings}{Section
3}, we detail the augmentation strategies implementation per encoding.

The proposed multi-encoding framework allows the training of our VAE
model using the same methods (i.e., encoding extraction, decoding of the
predictions, augmentation, one-hot encoding, storage, and retrieval of
the encoded dataset). The VAE model\textquotesingle s implementation,
loss functions, and measures rely on the Tensorflow\footnote{https://www.tensorflow.org,
  Last accessed on 07/03/2023.} framework. We detail its implementation
in \protect\hyperlink{vae-model}{Section
4}.

\hypertarget{symbolic-music-encodings}{%
\section{3. Symbolic Music Encodings}\label{symbolic-music-encodings}}

Departing from previous research on symbolic music encodings
(\protect\hyperlink{ref-https:ux2fux2fdoi.orgux2f10.48550ux2farxiv.2109.03454}{Prang
\& Esling, 2021};
\protect\hyperlink{ref-https:ux2fux2fdoi.orgux2f10.48550ux2farxiv.1709.01620}{Briot,
Hadjeres, \& Pachet, 2017}), we compare four popular corpus encodings
--- piano roll, MIDI, ABC, and Tonnetz --- in providing a pitch space
with optimal model reconstruction performance and pitch relations that
align with cognitive distances. Furthermore, we present two new
encodings, relying on the ability of the Fourier space to describe
musical objects and their intrinsic relations: the DFT of pitch class
distributions and the DFT of the piano roll. Sections
\protect\hyperlink{piano-roll}{3.1} to
\protect\hyperlink{dfts-of-pitch-and-pitch-class-distributions}{3.5}
present each encoding and its implementation within our work. We offer
an engaging platform\footnote{\url{https://nadiacarvalho.github.io/Latent-Tonal-Music/}}
for users to explore the encodings through various symbolic music
compositions.

\hypertarget{piano-roll}{%
\subsection{3.1. Piano roll}\label{piano-roll}}

\begin{figure}[h!]
\centering
\includegraphics[width=0.85\textwidth]{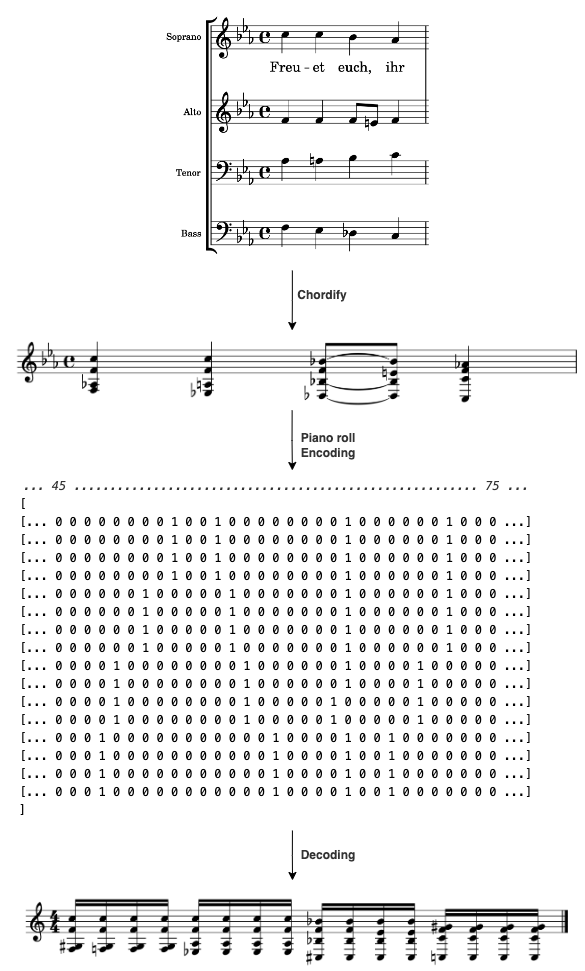}
\caption{Process of Encoding and Decoding the first measure of a "Freuet euch,
ihr Christen alle Bach" (BWV 40/8) as a piano roll (Original encoding,
0-128).}
\label{fig:figure-3}
\end{figure}

\begin{figure}[h!]
\centering
\includegraphics[width=1\textwidth]{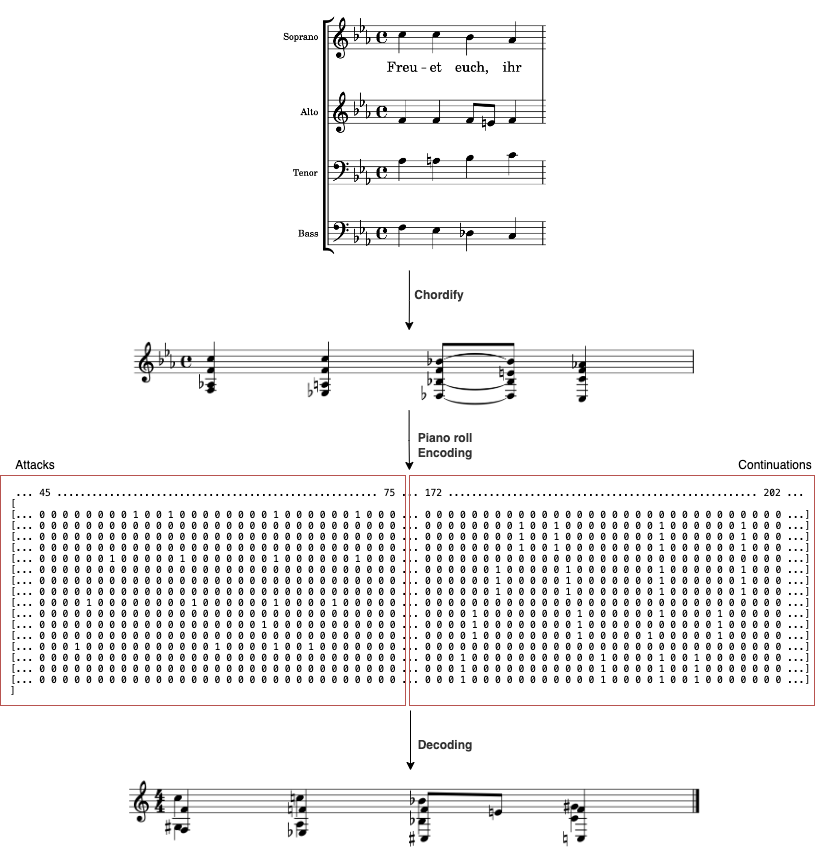}
\caption{Process of encoding and decoding the first measure of a "Freuet euch,
ihr Christen alle Bach" (BWV 40/8) as a piano roll. (Using our
``continuation''-based encoding, 0-128 for attack notes and 129-256 for
continuation notes).}
\label{fig:figure-4}
\end{figure}

The piano roll encoding uses a binary vector of ones and zeros
representing each note sequence\textquotesingle s timestep. Ones denote
note activation, and zeros represent the non-activated notes. This
method is widely used for encoding melodic and polyphonic music
structures and is known for its simplicity.~The most noteworthy
limitation, shown in Figure~\ref{fig:figure-3}, is its
inability to determine the end of each represented note. As shown in
Figure~\ref{fig:figure-4}, we address this limitation by
extending the piano roll encoding to twice its length. The first half of
the encoding pertains to notes starting at the timestep, while the
second half refers to active notes from previous timesteps, i.e.,
continuations.

To compute the augmentations, i.e., transposing the piano roll encoding
to a different key, we rotate the representation vector by the number of
half-tones corresponding to the transposing interval. This process is
done separately for attacks and continuations. First, we rotate the
first part of the vector containing the attacks. Second, we rotate the
second part containing the continuations to maintain the original
key\textquotesingle s encoding continuity.

\hypertarget{midi-like}{%
\subsection{3.2. MIDI-like}\label{midi-like}}

\begin{figure}[h!]
\centering
\includegraphics[width=0.95\textwidth]{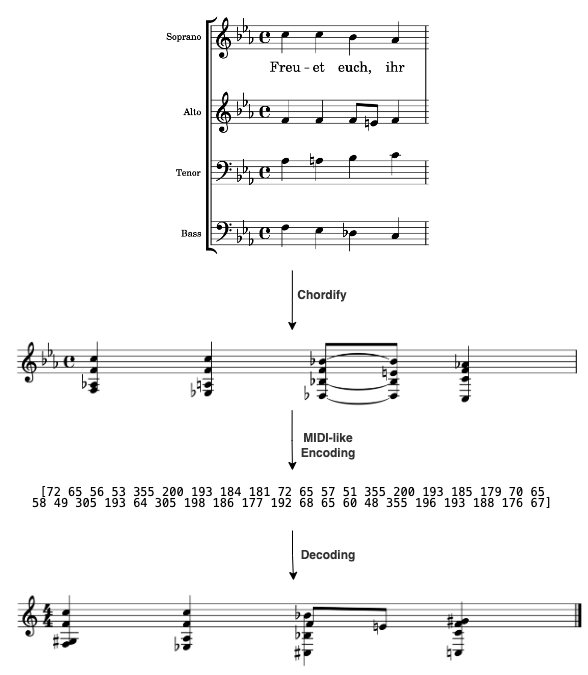}
\caption{Process of Encoding and Decoding the first measure of a "Freuet euch,
ihr Christen alle Bach" (BWV 40/8) with MIDI-like encoding.}
\label{fig:figure-5}
\end{figure}

Our work adopts the MIDI (Musical Instrument Digital Interface) protocol
as an encoding, as first proposed by
\protect\hyperlink{ref-temp_id_04921704639719682}{Oore et al. (2018)}.
This approach relies on a vocabulary of four main MIDI events, namely
the NOTE\_ON event, the corresponding NOTE\_OFF event, the SET\_VELOCITY
event, and the TIME\_SHIFT event. These events represent each timestep
of an input sequence as a discrete event, handling any form of music
with varying degrees of polyphony and metrical variation.
Figure~\ref{fig:figure-5} shows the extraction process
of the MIDI-like encoding.

MIDI-like encoding only has one value per timestep. Therefore, we can
perform augmentation by adding the number of half-tones corresponding to
the transposing interval to the MIDI-like encoded music. Similarly, when
the encoding is in its one-hot form,\footnote{One-hot encoding is a
  technique used to represent categorical data as binary vectors.} we
only need to rotate the encoded music horizontally by the same number of
elements as the transposing interval.

\hypertarget{abc}{%
\subsection{3.3. ABC}\label{abc}}

ABC notation uses letters, numbers, and symbols to represent musical
notes and rhythms. The system uses basic rules to represent each musical
element, such as pitch, duration, and ornamentation. One of the
advantages of ABC notation is its simplicity and ease of use, as it can
be quickly learned by musicians and non-musicians alike
(\protect\hyperlink{ref-temp_id_9840267697402025}{Sturm et al., 2018}).

\begin{figure}[h!]
\centering
\includegraphics[width=0.85\textwidth]{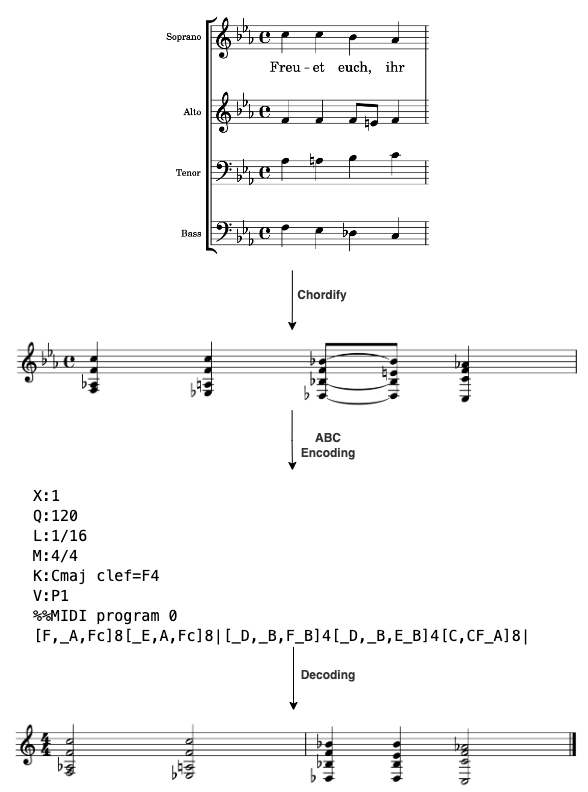}
\caption{Process of Encoding and Decoding the first measure of a "Freuet euch,
ihr Christen alle Bach" (BWV 40/8) with the ABC encoding.}
\label{fig:figure-6}
\end{figure}

ABC notation has been mainly applied to monophonic music structures,
i.e., melodies (\protect\hyperlink{ref-temp_id_10722088681995867}{Briot,
Hadjeres, \& Pachet, 2020}). We adopt it as a form of textual
representation of vertical or harmonic aggregates. To this end, we
developed a parser from the music21 structures to ABC.~Our parser
implementation is based on the Javascript Midi2ABC parser developed by
Marmoo.\footnote{Implementation is available at
  \url{https://github.com/marmooo/midi2abc,} last seen 07/03/2023.

  Online converter is available at
  \url{https://marmooo.github.io/midi2abc/,} last seen 07/03/2023.} The
process (see Figure~\ref{fig:figure-6}) involves
converting a segment of notes into ABC notation by dividing them by
instrument sections and adding headers for each instrument. The chords
and notes are then iterated through and converted to ABC notation,
taking into account component pitches, note duration, tie marks, rests,
and tuplets. The approach ensures correct timing and chord order.

We recognize some limitations in the ABC notation related to the
retainment of correct durations and passing notes, namely due to the
middle processing stage of chordification of the harmonic music texture,
as shown in Figure~\ref{fig:figure-6} (e.g., the F in
the second voice of the third chord would be lost). However, since the
study focuses on the cognitive distances from musical pitch within a
key, the former problem is not considered critical, as it ultimately may
discard some non-tonal tones or aggregate them into vertical slices. To
address the latter issue, we chose to incorporate passage notes by
splitting the tied notes of chords in which one voice moves, treating
them as distinct. This approach allowed us to retain more information
from the original music source in the encoded score, regardless of
possible absent notes.

\hypertarget{tonnetz}{%
\subsection{3.4. Tonnetz}\label{tonnetz}}

\protect\hyperlink{ref-10.5555ux2f3504035.3504298}{Chuan \& Herremans
(2018)} introduced an extended Tonnetz musical encoding based on the
Tonnetz graphical representation proposed by
\protect\hyperlink{ref-euler1739tentamen}{Euler (1739)}. Music theorists
and musicologists have long used the Tonnetz to investigate tonality and
tonal spaces.

\begin{figure}[h!]
\centering
\includegraphics[width=0.85\textwidth]{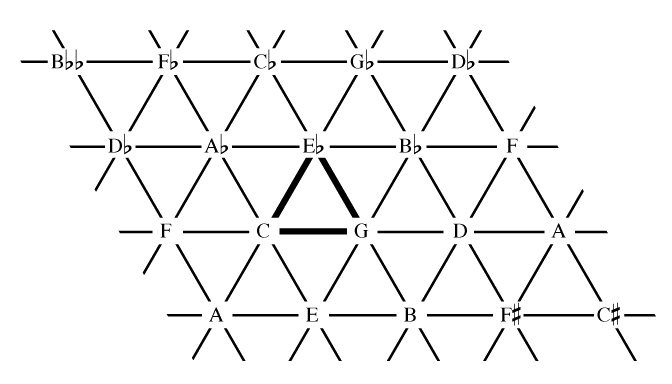}
\caption{Prototypical Tonnetz representation. Three nodes forming a triangle in
the network represent a triad, such as the ones in bold, representing
the C minus chord triad.}
\label{fig:figure-7}
\end{figure}

The construction of the Tonnetz typically involves using 12 pitch
classes, with nodes arranged in a circle-of-fifth sequence. Nodes to the
right create a cycle of perfect fifths, while nodes to the left form a
cycle of perfect fourths. Triangles in the network represent a triad,
with the parallel major and minor triads connected vertically by sharing
a baseline (see Figure~\ref{fig:figure-7}).

Our approach employs the expanded Tonnetz version proposed in
\protect\hyperlink{ref-10.5555ux2f3504035.3504298}{Chuan \& Herremans
(2018)}, utilizing a 24-by-12 matrix where each node represents a pitch
(not a pitch class, as in the traditional Tonnetz). The pitch register
information is maintained (from C0 to C\#8), determined by the proximity
to the central column\textquotesingle s pitch. Nodes on the same
horizontal line exhibit the circle-of-fifth relationship. The expansion
of the traditional one-octave Tonnetz facilitates simultaneous pitch
modeling, which is critical to encoding polyphonic music.

\begin{figure}[h!]
\centering
\includegraphics[width=1\textwidth]{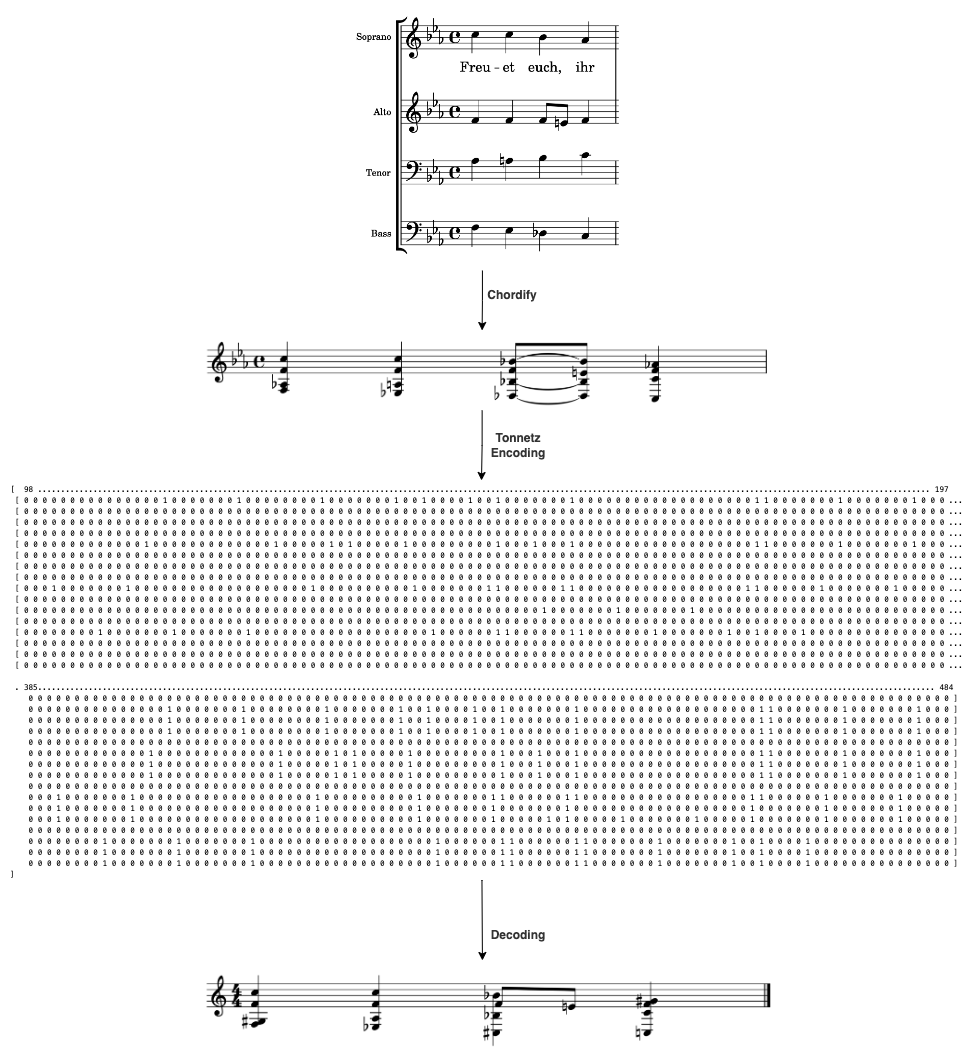}
\caption{Process of Encoding and Decoding the first measure of a "Freuet euch,
ihr Christen alle Bach" (BWV 40/8) with the Tonnetz encoding. For
legibility purposes, the continuations are below the attacks.
}
\label{fig:figure-8}
\end{figure}

We further extend the approach by duplicating the matrix horizontally to
capture attacks and continuations, using a piano roll-like approach
where active notes are encoded as ones and non-active notes as zeros
(see Figure~\ref{fig:figure-8}). However, as each pitch
appears at multiple positions in the matrix, all pitch positions must be
activated for each pitch. This makes computing Tonnetz augmentations
challenging, so transposed versions of symbolic music are encoded as new
Tonnetz encodings.

\hypertarget{dfts-of-pitch-and-pitch-class-distributions}{%
\subsection{3.5. DFTs of Pitch and Pitch Class
Distributions}\label{dfts-of-pitch-and-pitch-class-distributions}}

From previous work on adopting Fourier space to describe musical objects
and their intrinsic relations
(\protect\hyperlink{ref-temp_id_13915270107106203}{Quinn, 2006};
\protect\hyperlink{ref-temp_id_7430535358374943}{Yust, 2015};
\protect\hyperlink{ref-temp_id_851349609824851}{Amiot, 2016};
\protect\hyperlink{ref-temp_id_5576858097009489}{Bernardes et al.,
2016}), we adopt two different encodings based on the discrete Fourier
transform (DFT) of pitch distributions. The first applies the DFT on a
binary pitch distribution of \(m=128\) elements, similar to a piano
roll, i.e., a binary vector representation where active notes are
represented as ones. The second reduces such distribution to the 12
pitch classes, where\(m=12\). We adopt the non-trivial or
non-symmetrical output of the DFT, which results in \(m/2 + 1\) complex
numbers. The resulting vector can be converted into magnitude and phase
information from which musical objects can be interpreted. Magnitudes
encode the interval content of the represented pitch, and phases encode
the degree to which musical objects or vectors share common tones.

\begin{figure}[h!]
\centering
\includegraphics[width=.85\textwidth]{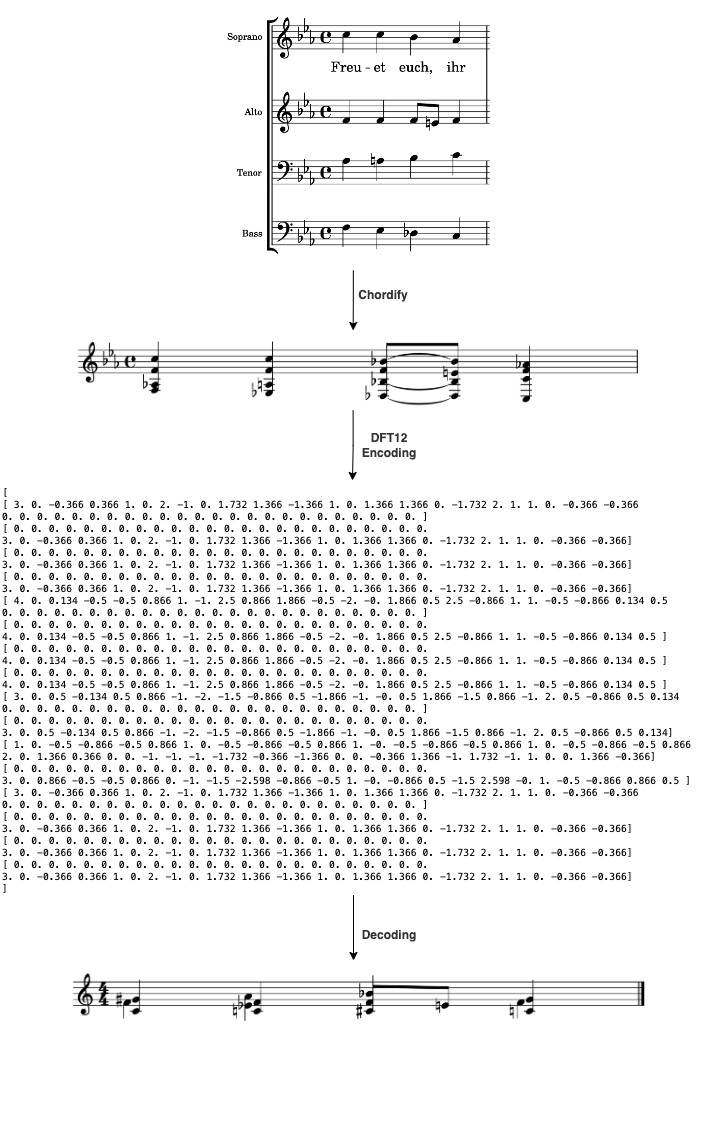}
\caption{Process of Encoding and Decoding the first measure of a "Freuet euch,
ihr Christen alle Bach" (BWV 40/8) with the encoding based on DFT from
Pitch Class Distribution.
}
\label{fig:figure-9}
\end{figure}

To represent a pitch class distribution using the Discrete Fourier
Transform (DFT), we first extract the pitch class information from the
notes and chords into a two-dimensional array of 24 columns per
timestep. Its position depends on whether it is attacked (first 12) or
continued. The resulting pitch information is transformed using the DFT
per note attacks and continuations (see
Figure~\ref{fig:figure-9}). The first component of the
list represents the number of attack activations, followed by a zero,
while the 25th component represents the number of continuation
activations, also followed by a zero.

\begin{figure}[h!]
\centering
\includegraphics[width=0.85\textwidth]{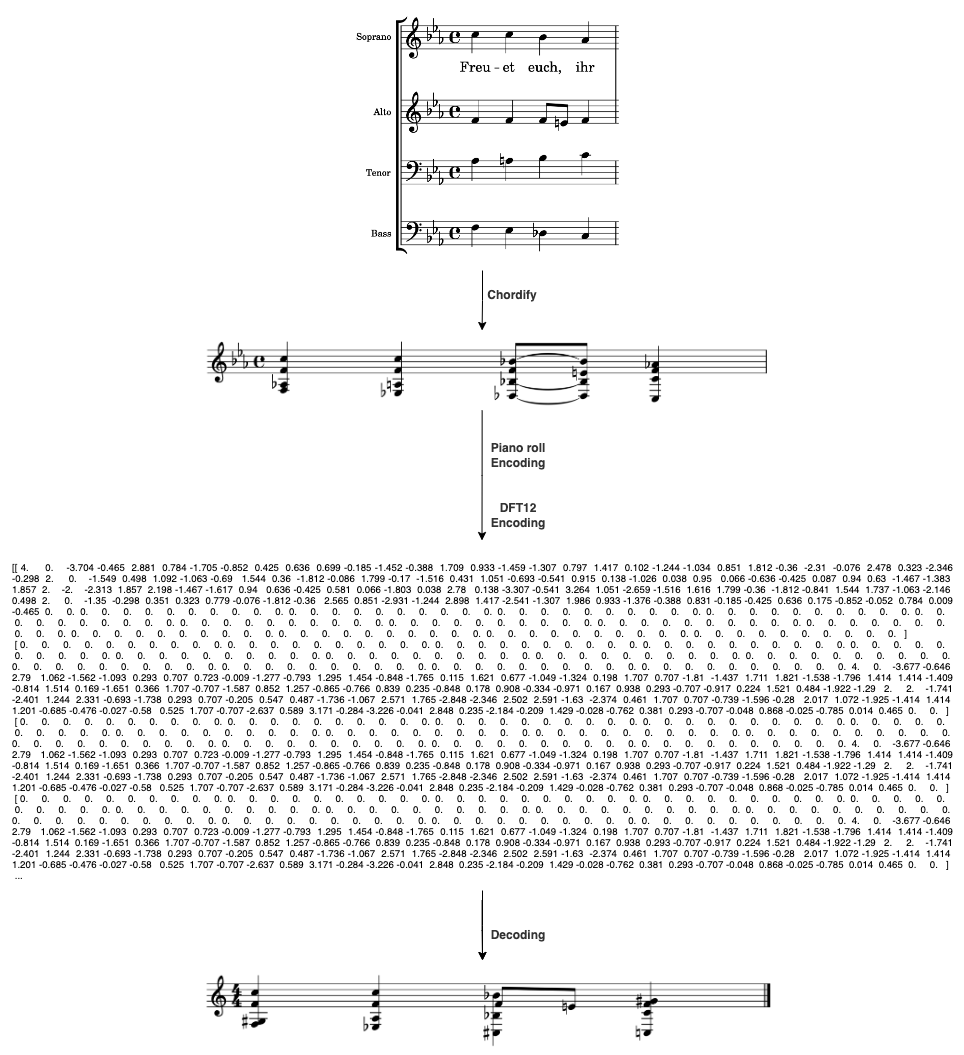}
\caption{Process of Encoding and Decoding the first measure of a "Freuet euch,
ihr Christen alle Bach" (BWV 40/8) with the encoding based on DFT from
Pitch. For legibility, we only show the output of the first four
timesteps.
}
\label{fig:figure-10}
\end{figure}

To perform the second DFT encoding, shown in Figure~\ref{fig:figure-10}, we use a similar process to
the first half. However, instead of creating a two-dimensional array
with 24 zeros per timestep, we compute a piano roll of 256 elements,
with 128 elements for attack activations and 128 for continuations. We
then apply the DFT separately on the attacks and continuations,
extracting only the non-symmetrical components, which are the first 64
components for each DFT output. Finally, we concatenate the two DFT
vectors into a unique representation.

We use the circular frequency shift property of the DFT to augment each
encoded timestep. Changing the pitch or pitch classes only affects the
phases of the DFT and not the magnitudes. To compute the new phases of
the encoding, we first compute the DFT of the pitch class (or pitch)
distribution with only the second component activated. For each complex
component of the output, we calculate and store its phase, using

\begin{equation}
a(z) = atan2(Im(z)/Re(z))
\label{eq:eq1}
\end{equation}

Then, to augment a specific song by a transposition \(t\), we rotate the
angles of the original DFT components (attacks and continuations are
processed individually), resulting in a new component with real and
imaginary parts, such that:

\begin{equation}
Re(z + t) = Re(z) * cos(t * a(z)) - Im(z) * sin(t * a(z))
\label{eq:eq2}
\end{equation}

\begin{equation}
Im(z + t) = Re(z) * sin(t * a(z)) + Im(z) * cos(t * a(z))
\label{eq:eq3}
\end{equation}

\vspace{1em}

\hypertarget{vae-model}{%
\section{4. VAE Model}\label{vae-model}}

We employ a VAE model
(\protect\hyperlink{ref-https:ux2fux2fdoi.orgux2f10.48550ux2farxiv.1312.6114}{Kingma
\& Welling, 2013}) to train and compare the latent space-generated
encodings. VAEs are generative neural network models that can learn a
compressed representation of data, such as images or music, by encoding
it into a low-dimensional latent space. VAEs are designed to generate
new data samples that resemble the input data by sampling from the
learned latent space
(\protect\hyperlink{ref-https:ux2fux2fdoi.orgux2f10.48550ux2farxiv.1312.6114}{Kingma
\& Welling, 2013}).

A VAE consists of two main components: 1) an \emph{encoder,
}\(e(x):\mathbb{R}^x, \mathbb{R}^z\), that maps the input data to a
lower-dimension latent space, and 2) a \emph{decoder,}
\(d(x):\mathbb{R}^z\mathbb{R}^x\), that maps the latent space back to
the input space, so that:~

\begin{equation}
\hat{x} = d(e(x)) \approx x 
\label{eq:eq4}
\end{equation}

During training, VAEs minimize a loss function that ensures the
generated data samples are similar to the input data while encouraging
the distribution of latent variables to follow a prior distribution,
such as a Gaussian distribution. This loss function is defined by two
parts (Equation~\ref{eq:eq5}): the reconstruction
loss, which evaluates how closely the output matches the input, and 2)
the KL-divergence loss, which quantifies the discrepancy between the
learned latent space and a prior distribution.

\begin{equation}
\log p_{\theta}(x)=\mathbb{E}_{q_{\phi}(z|x)}[\log [\frac{p_{\theta}(x,z)}{q_{\phi}(z|x)} ]] +D_{KL}(q_{\phi}(z|x) || p_{\theta}(z|x))
\label{eq:eq5}
\end{equation}

This encourages the model to generate diverse and realistic samples. In
the symbolic music domain, we note the widely known MusicVAE
(\protect\hyperlink{ref-https:ux2fux2fdoi.orgux2f10.48550ux2farxiv.1803.05428}{Roberts
et al., 2018}), which introduced a hierarchical decoder, the conductor,
that first outputs embeddings for subsequences of the input and then
uses these embeddings to generate each subsequence, independently.

\begin{figure}[h!]
\centering
\includegraphics[width=0.85\textwidth]{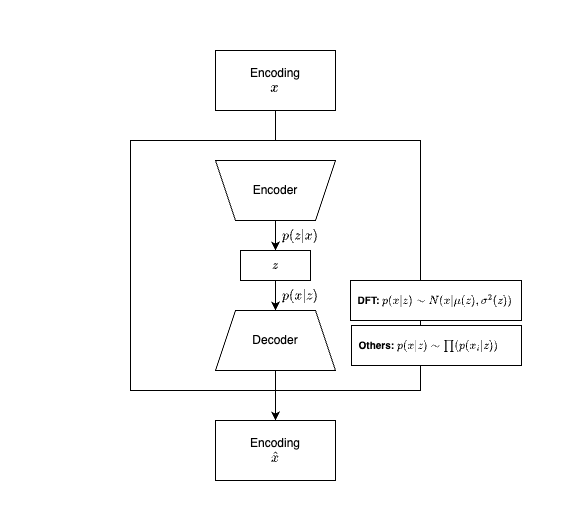}
\caption{Variational Encoding Implementation used in our experiments. Encoder and
Decoder are LSTM layers with 1024 units.
}
\label{fig:figure-11}
\end{figure}

To explore the correspondence between latent space structure and human
tonal perception, we opted for a flat baseline recurrent VAE model
instead of the more powerful MusicVAE, enabling a comparative study of
latent spaces and reconstruction possibilities across different
representations. Figure~\ref{fig:figure-11} shows the
implementation of the VAE. It consists of two recurrent LSTM layers with
1024 units, serving as encoder and decoder, respectively. We employ a
categorical cross-entropy reconstruction loss function for tokenized
one-hot encoded data (e.g., MIDI-like and ABC formats) and a binary
cross-entropy reconstruction loss function for multi-hot encoded data
(e.g., piano roll and Tonnetz formats). For DFT-encoded data, which
consists of float values, we use an MSE reconstruction loss function. To
minimize the loss function and optimize model parameters during
training, we use the Adam optimizer with an initial learning rate of
10E-4 and a batch size of 256. Finally, we employ a latent size of 256
for all musical encodings, which significantly reduces the input
dimensionality while preserving sufficient information for accurate
reconstruction
(\protect\hyperlink{ref-https:ux2fux2fdoi.orgux2f10.48550ux2farxiv.2109.03454}{Prang
\& Esling, 2021}). All VAE models were trained using these parameters.

\hypertarget{evaluation-results-and-discussion}{%
\section{5. Evaluation, Results, and
Discussion}\label{evaluation-results-and-discussion}}

\hypertarget{evaluation}{%
\subsection{\texorpdfstring{\textbf{5.1.
Evaluation}}{5.1. Evaluation}}\label{evaluation}}

In this section, we present a twofold evaluation strategy of the six
musical encodings described in
\protect\hyperlink{symbolic-music-encodings}{Section
3} trained in the VAE model defined in
\protect\hyperlink{vae-model}{Section 4} to
assess 1) the quality of musical embeddings in reconstructing data and
training the network and 2) the alignment of latent spaces to key
relations and distances in cognitive-led spaces. Ultimately, we aim to
assess the effectiveness of each musical encoding in training a VAE and
the ability of the latent space to capture tonal relations from the
input embeddings and define a data-driven pitch space. Our evaluation
and generated latent spaces adopt the JSB chorales as a test dataset,
extracted from the music21 library. The dataset is composed of 195
chorales in a major key (53\%) and 176 in a minor key (47\%), with G
Major (14\%), A minor (12\%), and G minor (11\%) being the most
representative keys. On average, a chorale is constructed of 84 chords.
We use 60\% of the chorales from the dataset for training and the other
40\% for testing. The training chorales are then augmented by
transposing (up and down) the encoding to the twelve keys.

\hypertarget{model-performance}{%
\subsubsection{\texorpdfstring{\textbf{5.1.1. Model
Performance}}{5.1.1. Model Performance}}\label{model-performance}}

To evaluate each encoding's ability to reconstruct the source data, we
examine the 1) accuracy, 2) MSE, and 3) KL-divergence scores for every
ten-timestep segment.~

Accuracy measures how well the reconstructed sequence matches the
original sequence and is reported in percentage. The highest accuracy
score of 100\% results from reconstructed sequences closely matching the
original sequences
(\protect\hyperlink{ref-temp_id_7415673136339067}{Briot \& Pachet,
2018}). MSE reckons the average squared difference between the
reconstructed and original sequences. A lower MSE score indicates that
the reconstructed sequence closely matches the original sequence
(\protect\hyperlink{ref-temp_id_7415673136339067}{Briot \& Pachet,
2018}). KL-divergence estimates the difference between two probability
distributions. In the context of sequence reconstruction, it measures
the difference between the original and reconstructed
sequence\textquotesingle s probability distributions. A lower
KL-divergence score, close to zero, indicates reconstructed sequences
closely matching the original sequence\textquotesingle s probability
distribution
(\protect\hyperlink{ref-https:ux2fux2fdoi.orgux2f10.48550ux2farxiv.1909.13668}{Prokhorov
et al., 2019}).

Additionally, we analyze the computational cost associated with training
each musical encoding on the VAE and provide insights into the
intelligibility and invariant procedure of the encodings.

\hypertarget{latent-space-analysis}{%
\subsubsection{\texorpdfstring{\textbf{5.1.2. Latent Space
Analysis~}}{5.1.2. Latent Space Analysis~}}\label{latent-space-analysis}}

\begin{figure}[h!]
\centering
\includegraphics[width=0.85\textwidth]{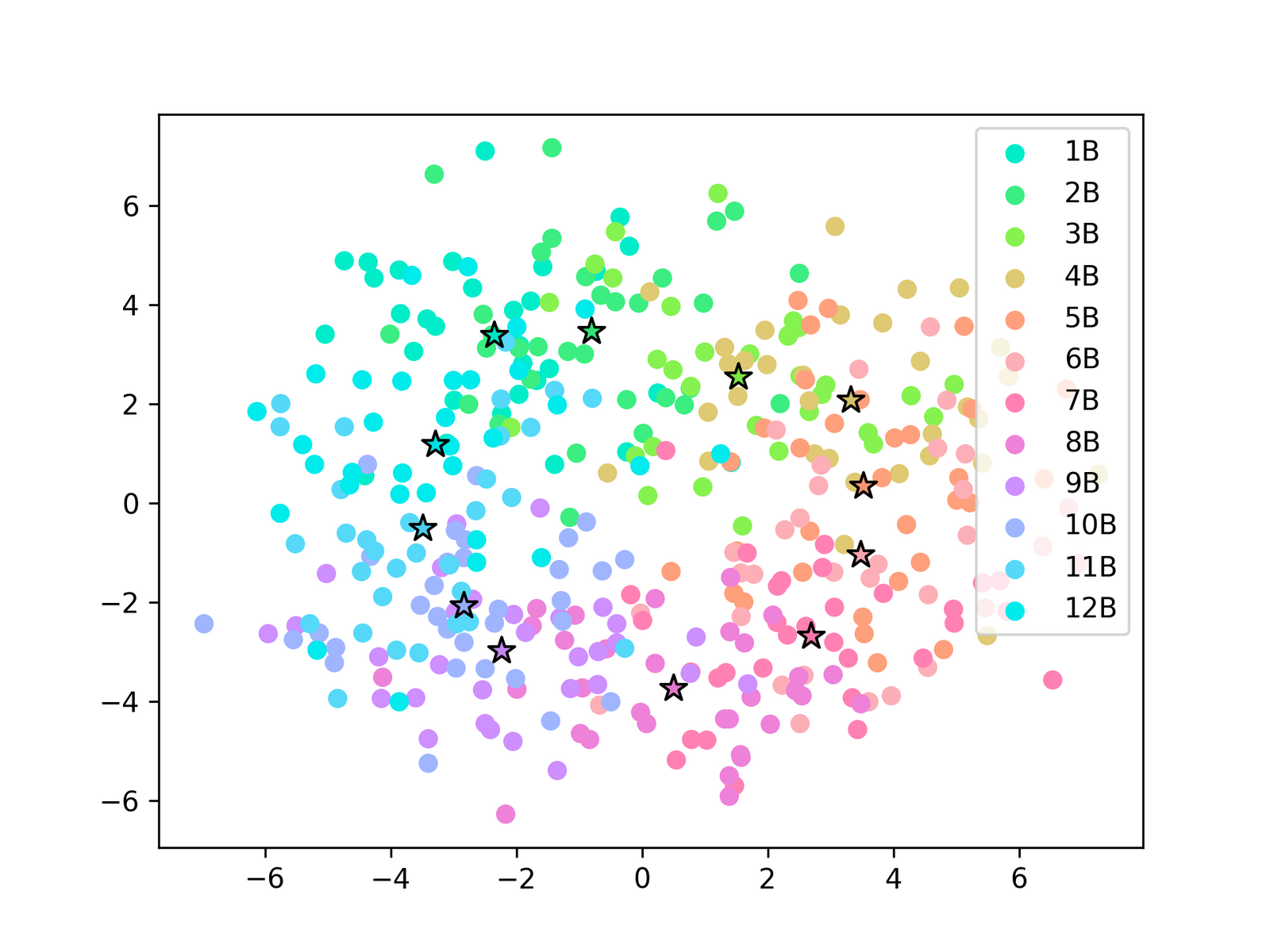}
\caption{Latent space of the choral ``Ich dank dir, lieber Herre'' (BWV 347) in A
major, transposed into the remaining 11 major keys from a DFT of pitch
distribution encoding. We adopt the Camelot Wheel for coloring and
numbering the clusters in the latent space (i.e., the keys B to E,
arranged by fifths, are represented by numbers 1 to 12, followed by a B
for the piece's major mode). Each key\textquotesingle s cluster centroid
is denoted by a colored star that matches the respective key.
}
\label{fig:figure-12}
\end{figure}

To inspect the alignment of the VAE latent space trained from different
encoding to cognitive-driven tonal space, we adopt a twofold strategy.
First, we project twelve annotated transpositions of a JSB chorale onto
the latent space, dividing them into ten-timestep segments. Second,
assuming each key is a cluster, we compute cluster metrics and
non-parametric circular statistics methods to evaluate its performance.
In other words, we aim to find how well each key transposition is
segmented and whether their spatial arrangement follows the circle of
fifths. Figure~\ref{fig:figure-12} shows the latent
space of the choral ``Ich dank dir, lieber Herre'' (BWV 347) in A major.
The choral is projected in all 12 major keys using a DFT of pitch
distribution encoding. When the originally projected choral is in the
minor mode, we transpose it to the remaining 11 minor keys.~While
plotting the latent space, we adopt the Camelot Wheel's\footnote{Davis,
  Mark, ``Harmonic Key Selection'', Camelot Sound,
  \url{http://www.camelotsound.com/Easymix.aspx.} Last accessed on
  07/03/2023.} colors and key colors and key enumerations (numbers 1-12
are keys B to E, by fifth intervals, while the following letter, A or B,
is the minor and major mode, respectively).

In detail, the key-annotated music segments from the encoded musical
data in all 12 keys are returned as a list of tuples. Then, we process
each segment's tuples through the pre-trained model\textquotesingle s
encoder to create its latent space.~To visualize and reduce the
computational complexity of the evaluation metrics, we employ principal
component analysis (PCA) to reduce the multidimensional latent space
data to two dimensions while retaining as much of the original data
variation as possible
(\protect\hyperlink{ref-temp_id_5755536052596777}{Pearson, 1901}).
Finally, we calculate their intra-segment distances per key and
inter-key distances from the resulting space to identify if they exist
as separate entities and somehow align with the expected fuzzy cluster
behavior from cognitive space. Within the tonal music context, the fuzzy
nature of the resulting clusters is expected due to the shared pitch
between keys. Two neighbor keys typically have one different pitch
element and share the remaining collection of pitches. Therefore, a
perfect key cluster separation is not expected. Furthermore, we compute
the order in which the keys are located in the space and thus their
relationships, which, as shown in Figure~\ref{fig:figure-1}, align with the circle of fifths.~ Distances in the latent space are
understood as the proximity or relation between two segments. Therefore,
the smaller the distances, the more related the two segments are
expected to be.

We adopt two cluster-evaluation metrics, namely the Davis-Bouldin score
(\protect\hyperlink{ref-temp_id_9132369059429659}{Davies and Bouldin,
1979}) and Dunn index
(\protect\hyperlink{ref-temp_id_47871824385491646}{Dunn†, 1974}), to
assess the intra-segment distances per key and inter-key distances from
the latent spaces. Each of the 12 key transpositions per dataset chorale
is understood here as a cluster. These measures roughly capture the
silhouette per key (i.e., intra-segment distances per key) and the
segregation between keys. In detail, the Davis-Bouldin score and Dunn
index capture the degree to which the resulting latent space provides a
compact and well-separated key cluster. The Davis-Bouldin score is lower
for better-separated clusters and higher for poorly separated clusters.
For the Dunn index, we adopt as intra-cluster metric (or cluster
diameter) the average Euclidean distance across all key cluster segments
and, as inter-cluster metric, the distance between each
cluster\textquotesingle s nearest neighbors. The Dunn Index score
results from the ratio between the inter-cluster and intra-cluster
distances. It aims to be maximized, indicating better separated and
compact clusters.

\begin{figure}[h!]
\centering
\includegraphics[width=1\textwidth]{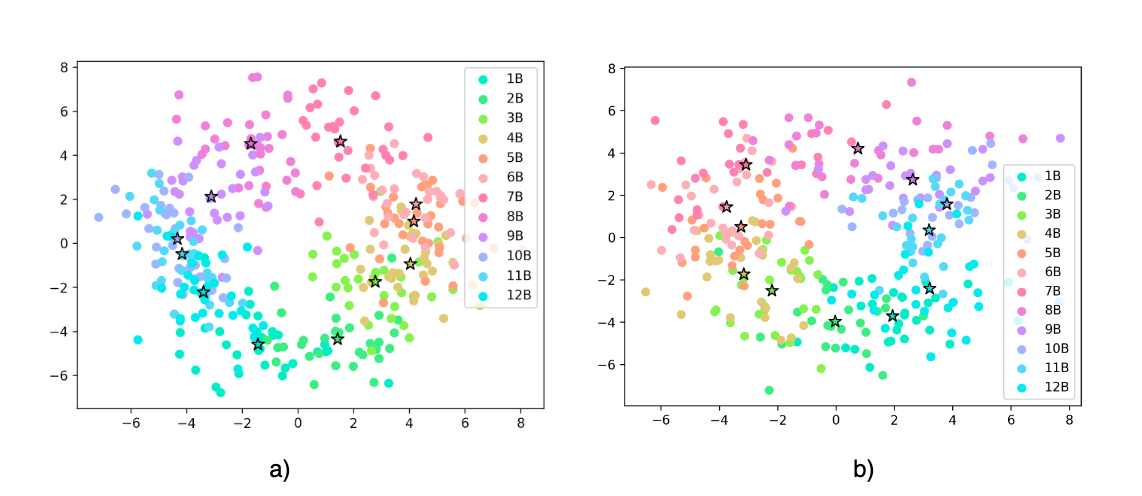}
\caption{Latent Spaces extracted from the Pitch DFT of two chorales in all twelve
key augmentations, whose Kendall's Tau coefficient is 1 (Fig. a) and -1
(Fig. b). The order of the circle of the fifths is exactly equal, but
the clusters are sequenced in either \textbf{anti-} or
\textbf{clockwise} direction. We adopt the Camelot Wheel for coloring
and numbering the clusters in the latent space (i.e., the keys B to E,
arranged by fifths, are represented by numbers 1 to 12, followed by a B
for the piece's major mode). Each key\textquotesingle s cluster centroid
is denoted by a colored star that matches the respective key.
}
\label{fig:figure-13}
\end{figure}

To assess the degree to which the key positions from latent space align
with the circle of fifth in cognitive spaces, we adopt a non-parametric
sample circular correlation coefficient measure. In detail, we apply the
circular non-parametric Kendall\textquotesingle s Tau, as proposed by
\protect\hyperlink{ref-temp_id_4403075070432403}{Fisher and Lee (1982)},
to measure the degree of association between a circular sequence of keys
in fifths and the resulting order of keys in each latent space. Tau
correlation coefficient ranges between -1 and 1, where -1 indicates a
perfect negative association, 0 indicates no association, and 1
indicates a perfect positive association between two variables. Optimal
correlations for our problem result from maximizing the absolute value
of the Tau coefficient, as both -1 and 1 capture the order of the keys
in circles of fifths (see Figure~\ref{fig:figure-13}).

\hypertarget{results-and-discussion}{%
\subsection{\texorpdfstring{\textbf{5.2. Results and
Discussion}}{5.2. Results and Discussion}}\label{results-and-discussion}}

\hypertarget{model-performance}{%
\subsubsection{\texorpdfstring{\textbf{5.2.1. Model
Performance}}{5.2.1. Model Performance}}\label{model-performance}}

Table~\ref{tab:table-1} presents objective results on
the model's reconstruction performance. Interestingly, KL-divergence and
MSE demonstrate consistent outcomes, while accuracy does not reflect the
same encoding order.

The encoding with the best accuracy in music reconstruction is ABC
notation (83\%), closely followed by MIDI-like (77\%), the two DFT
methods (77\% and 76\%). Both ABC and MIDI-like encodings, which use
categorical one-hot encoding, performed well in all three measures by
presenting the lowest values for both KL-divergence and MSE scores.
Therefore, they are expected to capture the most information from the
original chorales, aligning with the results of
\protect\hyperlink{ref-https:ux2fux2fdoi.orgux2f10.5281ux2fzenodo.4285410}{Plitsis
et al. (2020)}.

Interestingly, the Pitch Class Distribution DFT achieved a higher
accuracy score than the Pitch DFT, which suggests that the fewer
features of the latter may lead to better accuracy results. However, DFT
methods resulted in higher KL-divergence and MSE scores, indicating that
the reconstructed sequence and its probability distribution have poorer
matching when compared to the original sequence. At the same time, this
may be interpreted as a sign that these models are learning a more
informative latent space
(\protect\hyperlink{ref-https:ux2fux2fdoi.orgux2f10.48550ux2farxiv.1910.13181}{Ucar,
2019}).

\begin{table}[]
\centering
\begin{tabular}{p{1.8cm}ccccc}
\textbf{\begin{tabular}[c]{@{}c@{}}Musical \\Encoding\end{tabular}} &
  \textbf{\begin{tabular}[c]{@{}c@{}c@{}}Reconst.\\ Accuracy \\ (\%)\end{tabular}} &
  \textbf{\begin{tabular}[c]{@{}c@{}}KL\end{tabular}} &
  \textbf{MSE} &
  \textbf{\begin{tabular}[c]{@{}c@{}c@{}} Time per \\ Epoch \\(min)\footnote{Average
  Computation Time per Epoch computed on a Macbook-Pro with M1
  processor/GPU of 8 cores (2020).}\end{tabular}} &
  \textbf{\begin{tabular}[c]{@{}c@{}@{}}Embedding \\ Time \\ (min)\footnote{Average Computation Time per
  Epoch computed on a Macbook-Pro with M1 processor/GPU of 8 cores
  (2020).}\end{tabular}} \\ 
  \hline
\textbf{Piano roll}      & \textless 1 & 12.48  & 1.4E-02 & $\sim$50   & $\sim$2  \\
\textbf{MIDI-like}       & 77.4        & .80    & 8.0E-04 & $\sim$2h30 & $\sim$2  \\
\textbf{ABC}             & 82.6        & .54    & 6.7E-03 & $\sim$5h30 & $\sim$3  \\
\textbf{Tonnetz}         & \textless 1 & 35.29  & 1.9E-02 & $\sim$55   & $\sim$3  \\
\textbf{PC DFT} & 77.3        & 136.02 & 6.5E-01 & $\sim$1h10 & $\sim$5  \\
\textbf{P DFT}       & 75.6        & 472.79 & 9.1E-01 & $\sim$1h10 & $\sim$20
\end{tabular}
\caption{Results on the model’s reconstruction performance. PC DFT and P DFT stand for Pitch Class DFT and Pitch DFT.}
\label{tab:table-1}
\end{table}

The multi-hot encoded musical encodings (piano roll and tonnetz) have
the poorest model performance, as reflected by the remarkably low
accuracy scores. Moreover, these representations are still prone to
overfitting despite our effort to minimize it through the regularisation
techniques, such as dropout and data augmentation.

In terms of computational cost, the piano roll had the shortest training
time, closely followed by the tonnetz encoding. The DFT encodings had
similar training times, although not as low as the previous two. In
contrast, the ABC and MIDI-like encodings require much more sequences to
train than the others, even with the same sequence length. Consequently,
they require significantly longer training times. However, both models
converge in higher values of accuracy (and lower KL-divergence and MSE)
when compared to the remaining encodings. Notably, training the musical
embedding from ABC notation requires a longer time per epoch compared to
the others.

During the process of extracting information from the original symbolic
music, the Pitch DFT exhibits the slowest performance by a substantial
margin, taking almost four times longer than the second-worst performing
method, the DFT of pitch class. In contrast, the remaining musical
encodings require nearly equal amounts of time for feature extraction,
with the piano roll achieving the best performance by less than five
seconds.

Additionally, we observed that augmenting the piano roll and MIDI-like
embeddings is a simple and fast task. However, the process of augmenting
the DFTs, Tonnetz, and ABC encodings is slow, particularly for the
latter two. To address this, we pre-computed and saved the augmentations
prior to training, allowing us to load them during the training
preparation phase.

\hypertarget{latent-space-analysis}{%
\subsubsection{\texorpdfstring{\textbf{5.2.2. Latent Space
Analysis~}}{5.2.2. Latent Space Analysis~}}\label{latent-space-analysis}}

Table~\ref{tab:table-2} presents the average and
standard deviation values for the cluster and key-distance metrics,
allowing us to evaluate the alignment between the latent space of each
musical encoding and cognitive spaces.

\begin{table}[]
\begin{tabular}{p{2em}lccc}
\multicolumn{2}{l}{\multirow{3}{*}{\textbf{\begin{tabular}[c]{@{}c@{}}Musical \\Encoding\end{tabular}}}} &
  \multirow{3}{*}{\textbf{\begin{tabular}[c]{@{}c@{}}Davis-Bouldin\\ Score\end{tabular}}} &
  \multirow{3}{*}{\textbf{\begin{tabular}[c]{@{}c@{}}Dunn\\ Index\end{tabular}}} &
  \multirow{3}{*}{\textbf{\begin{tabular}[c]{@{}c@{}}Kendall's\\ Tau\end{tabular}}} \\
\multicolumn{2}{c}{}                         &             &                 &           \\
\multicolumn{2}{c}{}                         &             &                 &           \\ \hline
\multicolumn{2}{l}{\textbf{Piano roll}}      & 32.8 ± 17.8 & .0005 ± .0005   & .11 ± .08 \\
\multicolumn{2}{l}{\textbf{MIDI-like}}       & 55.9 ± 35.3 & .0001 ± .0001   & .15 ± .14 \\
\multicolumn{2}{l}{\textbf{ABC}}             & 66.9 ± 55.0 & .00003 ± .00003 & .11 ± .09 \\
\multicolumn{2}{l}{\textbf{Tonnetz}}         & 33.0 ± 21.6 & .0006 ± .0007   & .11 ± .08 \\
\multicolumn{2}{l}{\textbf{Pitch Class DFT}} & 37.3 ± 59.1 & .0006 ± .0006   & .11 ± .08 \\
\multicolumn{2}{l}{\textbf{Pitch DFT}}       & 8.1 ± 5.5   & .0008 ± .0009   & .44 ± .32
\end{tabular}
\caption{Average and standard deviation values for the cluster and key-distance metrics}\label{tab:table-2}
\end{table}

The latent space trained on Pitch DFT encoding demonstrates the best
alignment with cognitive spaces, surpassing other encodings in all three
metrics, as anticipated due to its high accuracy and KL-divergence
values, and in line with the preference for signal-based encoding
concluded in
\protect\hyperlink{ref-https:ux2fux2fdoi.orgux2f10.48550ux2farxiv.2109.03454}{Prang
\& Esling, (2021)}. It presents the most condensed and seamless
representation in the latent space, outperforming other encodings by a
large margin. The piano roll, Tonnetz, and Pitch Class DFT encodings
also show relatively good results, while the MIDI-like and ABC encodings
have the worst scores in all three metrics, indicating poorer clustering
performance.

Upon analyzing Pitch DFT, there are several noteworthy insights. First,
around 16\% of chorales display Kendall Tau\textquotesingle s absolute
values exceeding .9, indicating a strong alignment with the cognitive
pitch space. Moreover, major key chorales\textquotesingle{} latent
spaces appear to be more efficient in capturing these distances, with
87\% of chorales that have Kendall Tau\textquotesingle s absolute values
greater than .9 being in a major key. Surprisingly, the findings also
reveal that longer chorales are better at capturing these distances.
About 30\% of chorales containing over 114 slices show Tau absolute
values greater than .9, in contrast to only 10\% of chorales with less
than 64 slices exhibiting such values. Based on conventional
assumptions, we would anticipate that longer chorales, which are more
susceptible to modulations, would have latent spaces that are more
ambiguous and, therefore, less aligned with the cognitive pitch space.

\hypertarget{conclusions-and-future-work}{%
\section{6. Conclusions and Future
Work}\label{conclusions-and-future-work}}

Our paper explores the performance of VAEs in reconstructing tonal
symbolic music and eliciting latent representations of cognitive and
musical theoretical value. We trained VAEs on a prototypical tonal music
corpus of 371 Bach\textquotesingle s chorales, represented as six
different symbolic music encodings (i.e., Piano roll, MIDI, ABC,
Tonnetz, DFT of pitch and pitch class distributions) and evaluated the
degree to which the latent spaces defined by the different VAE corpus
encodings align with cognitive distances from musical pitch, based on
objective reconstruction performance metrics (accuracy, MSE, and
KL-divergence), computational performance, and clustering metrics
(Davis-Bouldin Score, Dunn Index, and Kendall\textquotesingle s Tau).
Our VAE implementation and the encodings framework are available online
at \url{https://github.com/NadiaCarvalho/Latent-Tonal-Music}. Furthermore, on our website \url{https://nadiacarvalho.github.io/Latent-Tonal-Music/}, we offer an engaging platform for users to explore the encodings through various symbolic music compositions.

The results showed that the ABC VAE performed best in the data
reconstruction performance metrics, while the proposed Pitch DFT VAE
latent space is better aligned with a common-tone space where
overlapping objects within a key are fuzzy clusters, which impose a
well-defined order of structural significance or stability, i.e., a
tonal hierarchy. In sum, ABC encodings would be preferable when there is
an interest in preserving the original symbolic musical structures,
while Pitch DFT VAEs can produce more diverse and varied generative
models. Moving forward, we plan to conduct a more in-depth analysis of
the data reconstructed using this encoding.

The findings suggest potential for exploring pitch spaces in less
structured harmonic or pitch systems, such as modal and microtonal
music. While many existing pitch spaces accurately represent the
distances across various hierarchies (e.g., pitches, chords, and keys),
there are currently no such spaces available for non-tonal music
expressions, as far as we know.

\hypertarget{acknowledgments}{%
\section{\texorpdfstring{\textbf{Acknowledgments}}{Acknowledgments}}\label{acknowledgments}}

This research has been funded by the Portuguese National Funding Agency
for Science, Research and Technology {[}2021.05132.BD{]}.

\hypertarget{ethics-statement}{%
\section{\texorpdfstring{\textbf{Ethics
Statement}}{Ethics Statement}}\label{ethics-statement}}

This research study did not involve human subjects, animals, or
sensitive data. Therefore, no ethics approval was required. The authors
have no conflicts of interest to declare.

\hypertarget{refs}{%
\section{\texorpdfstring{\textbf{References}}{References}}\label{references}}

\begin{CSLReferences}{1}{0}
\leavevmode\vadjust pre{\hypertarget{ref-temp_id_851349609824851}{}}%
Amiot, Emmanuel. 2016. \emph{Music Through Fourier Space}.
\emph{Computational Music Science}. Springer International Publishing.
\url{https://doi.org/10.1007/978-3-319-45581-5}.

\leavevmode\vadjust pre{\hypertarget{ref-temp_id_5576858097009489}{}}%
Bernardes, Gilberto, Diogo Cocharro, Marcelo Caetano, Carlos Guedes, and
Matthew E.P. Davies. 2016. {``A Multi-Level Tonal Interval Space for
Modelling Pitch Relatedness and Musical Consonance.''} \emph{Journal of
New Music Research} 45 (4): 281--94.
\url{https://doi.org/10.1080/09298215.2016.1182192}.

\leavevmode\vadjust pre{\hypertarget{ref-https:ux2fux2fdoi.orgux2f10.48550ux2farxiv.1709.01620}{}}%
Briot, Jean-Pierre, Gaëtan Hadjeres, and François-David Pachet. 2017.
{``Deep Learning Techniques for Music Generation -\/- A Survey.''}
arXiv. \url{https://doi.org/10.48550/ARXIV.1709.01620}.

\leavevmode\vadjust pre{\hypertarget{ref-temp_id_10722088681995867}{}}%
Briot, Jean-Pierre, Gaëtan Hadjeres, and François-David Pachet. 2020. \emph{Deep Learning Techniques for Music Generation}.
\emph{Computational Synthesis and Creative Systems}. Springer
International Publishing.
\url{https://doi.org/10.1007/978-3-319-70163-9}.

\leavevmode\vadjust pre{\hypertarget{ref-temp_id_7415673136339067}{}}%
Briot, Jean-Pierre, and François Pachet. 2018. {``Deep Learning for
Music Generation: Challenges and Directions.''} \emph{Neural Computing
and Applications} 32 (4): 981--93.
\url{https://doi.org/10.1007/s00521-018-3813-6}.

\leavevmode\vadjust pre{\hypertarget{ref-bryan-kinns2021exploring}{}}%
Bryan-Kinns, Nick, Berker Banar, Corey Ford, Courtney N. Reed, Yixiao
Zhang, Simon Colton, and Jack Armitage. 2021. {``Exploring XAI for the
Arts: Explaining Latent Space in Generative Music.''} In
\emph{eXplainable AI Approaches for Debugging and Diagnosis.}
\url{https://openreview.net/forum?id=GLhY_0xMLZr}.

\leavevmode\vadjust pre{\hypertarget{ref-10.5555ux2f3504035.3504298}{}}%
Chuan, Ching-Hua, and Dorien Herremans. 2018. {``Modeling Temporal Tonal
Relations in Polyphonic Music Through Deep Networks with a Novel
Image-Based Representation.''} In \emph{Proceedings of the Thirty-Second
AAAI Conference on Artificial Intelligence and Thirtieth Innovative
Applications of Artificial Intelligence Conference and Eighth AAAI
Symposium on Educational Advances in Artificial Intelligence}.
AAAI'18/IAAI'18/EAAI'18. New Orleans, Louisiana, USA: AAAI Press.

\leavevmode\vadjust pre{\hypertarget{ref-temp_id_043057691023887346}{}}%
Cohn, Richard. 1997. {``Neo-Riemannian Operations, Parsimonious
Trichords, and Their {`Tonnetz'} Representations.''} \emph{Journal of
Music Theory} 41 (1): 1. \url{https://doi.org/10.2307/843761}.

\leavevmode\vadjust pre{\hypertarget{ref-temp_id_22337014670269895}{}}%
Cohn, Richard. 1998. {``Introduction to Neo-Riemannian Theory: A Survey and
a Historical Perspective.''} \emph{Journal of Music Theory} 42 (2): 167.
\url{https://doi.org/10.2307/843871}.

\leavevmode\vadjust pre{\hypertarget{ref-temp_id_9132369059429659}{}}%
Davies, David L., and Donald W. Bouldin. 1979. {``A Cluster Separation
Measure.''} \emph{IEEE Transactions on Pattern Analysis and Machine
Intelligence} PAMI-1 (2): 224--27.
\url{https://doi.org/10.1109/tpami.1979.4766909}.

\leavevmode\vadjust pre{\hypertarget{ref-temp_id_6497209108989883}{}}%
Deutsch, Diana. 1984. {``Two Issues Concerning Tonal Hierarchies:
Comment on Castellano, Bharucha, and Krumhansl.''} \emph{Journal of
Experimental Psychology: General} 113 (3): 413--16.
\url{https://doi.org/10.1037/0096-3445.113.3.413}.

\leavevmode\vadjust pre{\hypertarget{ref-temp_id_47871824385491646}{}}%
Dunn†, J. C. 1974. {``Well-Separated Clusters and Optimal Fuzzy
Partitions.''} \emph{Journal of Cybernetics} 4 (1): 95--104.
\url{https://doi.org/10.1080/01969727408546059}.

\leavevmode\vadjust pre{\hypertarget{ref-euler1739tentamen}{}}%
Euler, Leonhard. 1739. \emph{Tentamen Novae Theoriae Musicae: Ex
Certissimis Harmoniae Principiis Dilucide Expositae}. ex typographia
Academiae scientiarum.

\leavevmode\vadjust pre{\hypertarget{ref-temp_id_4403075070432403}{}}%
Fisher, N. I., and A. J. Lee. 1982. {``Nonparametric Measures of
Angular-Angular Association.''} \emph{Biometrika} 69 (2): 315.
\url{https://doi.org/10.2307/2335405}.

\leavevmode\vadjust pre{\hypertarget{ref-https:ux2fux2fdoi.orgux2f10.48550ux2farxiv.2212.00973}{}}%
Guo, Z., J. Kang, and D. Herremans. 2022. {``A Domain-Knowledge-Inspired
Music Embedding Space and a Novel Attention Mechanism for Symbolic Music
Modeling.''} arXiv. \url{https://doi.org/10.48550/ARXIV.2212.00973}.

\leavevmode\vadjust pre{\hypertarget{ref-temp_id_9699009762824167}{}}%
Kim, Seung-Goo. 2022. {``On the Encoding of Natural Music in
Computational Models and Human Brains.''} \emph{Frontiers in
Neuroscience} 16 (September).
\url{https://doi.org/10.3389/fnins.2022.928841}.

\leavevmode\vadjust pre{\hypertarget{ref-https:ux2fux2fdoi.orgux2f10.48550ux2farxiv.1312.6114}{}}%
Kingma, Diederik P, and Max Welling. 2013. {``Auto-Encoding Variational
Bayes.''} arXiv. \url{https://doi.org/10.48550/ARXIV.1312.6114}.

\leavevmode\vadjust pre{\hypertarget{ref-krumhansl:1990}{}}%
Krumhansl, Carol L. 1990. \emph{Cognitive Foundations of Musical Pitch}.
Oxford Psychology Series, no. 17. New York: Oxford University Press.

\leavevmode\vadjust pre{\hypertarget{ref-lewin2010generalized}{}}%
Lewin, David. 1987. \emph{Generalized Musical Intervals and
Transformations}. New Haven: Yale University Press.

\leavevmode\vadjust pre{\hypertarget{ref-longuet:87}{}}%
Longuet-Higgins, H. C. 1987. \emph{Mental Processes: Studies in
Cognitive Science}. Explorations in Cognitive Science 1. Cambridge,
Mass: MIT Press.

\leavevmode\vadjust pre{\hypertarget{ref-temp_id_7741585219955907}{}}%
Mezza, Alessandro Ilic, Massimiliano Zanoni, and Augusto Sarti. 2023.
{``A Latent Rhythm Complexity Model for Attribute-Controlled Drum
Pattern Generation.''} \emph{EURASIP Journal on Audio, Speech, and Music
Processing} 2023 (1). \url{https://doi.org/10.1186/s13636-022-00267-2}.

\leavevmode\vadjust pre{\hypertarget{ref-temp_id_4779845625814849}{}}%
Moss, Fabian C., Markus Neuwirth, and Martin Rohrmeier. 2022. {``The
Line of Fifths and the Co-Evolution of Tonal Pitch-Classes.''}
\emph{Journal of Mathematics and Music} 17 (2): 173--97.
\url{https://doi.org/10.1080/17459737.2022.2044927}.

\leavevmode\vadjust pre{\hypertarget{ref-RePEc:wsi:acsxxx:v:25:y:2022:i:05n06:n:s0219525922400082}{}}%
Nardelli, Marco Buongiorno, Garland Culbreth, and Miguel Fuentes. 2022.
{``Towards A Measure Of Harmonic Complexity In Western Classical
Music.''} \emph{Advances in Complex Systems (ACS)} 25 (05n06): 1--11.
\url{https://doi.org/10.1142/S0219525922400082}.

\leavevmode\vadjust pre{\hypertarget{ref-temp_id_04921704639719682}{}}%
Oore, Sageev, Ian Simon, Sander Dieleman, Douglas Eck, and Karen
Simonyan. 2018. {``This Time with Feeling: Learning Expressive Musical
Performance.''} \emph{Neural Computing and Applications} 32 (4):
955--67. \url{https://doi.org/10.1007/s00521-018-3758-9}.

\leavevmode\vadjust pre{\hypertarget{ref-temp_id_5755536052596777}{}}%
Pearson, Karl. 1901. {``LIII. \emph{On Lines and Planes of Closest Fit
to Systems of Points in Space}.''} \emph{The London, Edinburgh, and
Dublin Philosophical Magazine and Journal of Science} 2 (11): 559--72.
\url{https://doi.org/10.1080/14786440109462720}.

\leavevmode\vadjust pre{\hypertarget{ref-https:ux2fux2fdoi.orgux2f10.5281ux2fzenodo.4285410}{}}%
Plitsis, Manos, Kosmas Kritsis, Maximos Kaliakatsos-Papakostas, Aggelos
Pikrakis, and Vassilis Katsouros. 2020. {``Towards a Classification and
Evaluation of Symbolic Music Encodings for RNN Music Generation.''}
\emph{Zenodo}, October. \url{https://doi.org/10.5281/ZENODO.4285410}.

\leavevmode\vadjust pre{\hypertarget{ref-https:ux2fux2fdoi.orgux2f10.48550ux2farxiv.2109.03454}{}}%
Prang, Mathieu, and Philippe Esling. 2021. {``Signal-Domain
Representation of Symbolic Music for Learning Embedding Spaces.''}
\emph{arXiv}. \url{https://doi.org/10.48550/ARXIV.2109.03454}.

\leavevmode\vadjust pre{\hypertarget{ref-https:ux2fux2fdoi.orgux2f10.48550ux2farxiv.1909.13668}{}}%
Prokhorov, Victor, Ehsan Shareghi, Yingzhen Li, Mohammad Taher Pilehvar,
and Nigel Collier. 2019. {``On the Importance of the Kullback-Leibler
Divergence Term in Variational Autoencoders for Text Generation.''}
arXiv. \url{https://doi.org/10.48550/ARXIV.1909.13668}.

\leavevmode\vadjust pre{\hypertarget{ref-temp_id_08049428684638582}{}}%
Qiu, Lvyang, Shuyu Li, and Yunsick Sung. 2021. {``3D-DCDAE: Unsupervised
Music Latent Representations Learning Method Based on a Deep 3D
Convolutional Denoising Autoencoder for Music Genre Classification.''}
\emph{Mathematics} 9 (18): 2274.
\url{https://doi.org/10.3390/math9182274}.

\leavevmode\vadjust pre{\hypertarget{ref-temp_id_13915270107106203}{}}%
Quinn, Ian. 2006. {``General Equal-Tempered Harmony (Introduction and
Part I).''} \emph{Perspectives of New Music} 44 (2): 114--58.
\url{https://doi.org/10.1353/pnm.2006.0010}.

\leavevmode\vadjust pre{\hypertarget{ref-https:ux2fux2fdoi.orgux2f10.48550ux2farxiv.1803.05428}{}}%
Roberts, Adam, Jesse Engel, Colin Raffel, Curtis Hawthorne, and Douglas
Eck. 2018. {``A Hierarchical Latent Vector Model for Learning Long-Term
Structure in Music.''} \emph{arXiv}.
\url{https://doi.org/10.48550/ARXIV.1803.05428}.

\leavevmode\vadjust pre{\hypertarget{ref-https:ux2fux2fdoi.orgux2f10.48550ux2farxiv.2302.05393}{}}%
Sarmento, Pedro, Adarsh Kumar, Yu-Hua Chen, CJ Carr, Zack Zukowski, and
Mathieu Barthet. 2023. {``GTR-CTRL: Instrument and Genre Conditioning
for Guitar-Focused Music Generation with Transformers.''} \emph{arXiv}.
\url{https://doi.org/10.48550/ARXIV.2302.05393}.

\leavevmode\vadjust pre{\hypertarget{ref-temp_id_38492172090114884}{}}%
Shepard, Roger N. 1982. {``Geometrical Approximations to the Structure
of Musical Pitch.''} \emph{Psychological Review} 89 (4): 305--33.
\url{https://doi.org/10.1037/0033-295x.89.4.305}.

\leavevmode\vadjust pre{\hypertarget{ref-temp_id_9840267697402025}{}}%
Sturm, Bob L., Oded Ben-Tal, Úna Monaghan, Nick Collins, Dorien
Herremans, Elaine Chew, Gaëtan Hadjeres, Emmanuel Deruty, and François
Pachet. 2018. {``Machine Learning Research That Matters for Music
Creation: A Case Study.''} \emph{Journal of New Music Research} 48 (1):
36--55. \url{https://doi.org/10.1080/09298215.2018.1515233}.

\leavevmode\vadjust pre{\hypertarget{ref-temp_id_7657716331815574}{}}%
Turker, Meliksah, Alara Dirik, and Pinar Yanardag. 2022. {``MIDISpace:
Finding Linear Directions in Latent Space for Music Generation.''} In
\emph{Creativity and Cognition}. ACM.
\url{https://doi.org/10.1145/3527927.3532790}.

\leavevmode\vadjust pre{\hypertarget{ref-tymoczko:10}{}}%
Tymoczko, Dmitri. 2010. \emph{A Geometry of Music: Harmony and
Counterpoint in the Extended Common Practice}. Oxford Studies in Music
Theory. New York: Oxford University Press.

\leavevmode\vadjust pre{\hypertarget{ref-https:ux2fux2fdoi.orgux2f10.48550ux2farxiv.1910.13181}{}}%
Ucar, Talip. 2019. {``Bridging the ELBO and MMD.''} arXiv.
\url{https://doi.org/10.48550/ARXIV.1910.13181}.

\leavevmode\vadjust pre{\hypertarget{ref-weber1832versuch}{}}%
Weber, Gottfried. 1817-1821. \emph{Versuch Einer Geordneten Theorie Der
Tonsetzkunst}. Vol. 1. B. Schott's Söhne.

\leavevmode\vadjust pre{\hypertarget{ref-temp_id_7430535358374943}{}}%
Yust, Jason. 2015. {``Applications of DFT to the Theory of
Twentieth-Century Harmony.''} In \emph{Mathematics and Computation in
Music}, 207--18. Springer International Publishing.
\url{https://doi.org/10.1007/978-3-319-20603-5_22}.

\end{CSLReferences}

\end{document}